%% file: pub_arxiv.tex
\documentclass[aps,prl,twocolumn,superscriptaddress,showpacs,preprintnumbers,amsmath,amssymb]{revtex4}
\usepackage{graphicx} 
\usepackage{dcolumn}  
\usepackage{array}
\usepackage{color}

\graphicspath{{ps}}

\begin{document}



\title{ \quad\\[1.0cm] Search for the decay {\boldmath $B^+\rightarrow\overline{K}{}^{*0}K^{*+}$} at Belle}

\input{author}

\begin{abstract}
We report a search for the rare charmless decay $B^+\rightarrow\overline{K}{}^{*0}K^{*+}$ using a data sample of $772\times10^6$ $B\bar{B}$ pairs collected at the $\Upsilon(4S)$ resonance with the Belle detector at the KEKB asymmetric-energy $e^+e^-$ collider. No statistically significant signal is found and a 90\% confidence-level upper limit is set on the decay branching fraction as $ \mathcal{B}(B^+\rightarrow\overline{K}{}^{*0}K^{*+}) <1.31\times 10^{-6}$.
\end{abstract}


\maketitle

\tighten

{\renewcommand{\thefootnote}{\fnsymbol{footnote}}}
\setcounter{footnote}{0}
%
%

  \begin{figure}[tb]
    \centering
    \vspace{-0.5cm}
    \hspace{-0.5cm}
    \includegraphics[width=4.7cm]{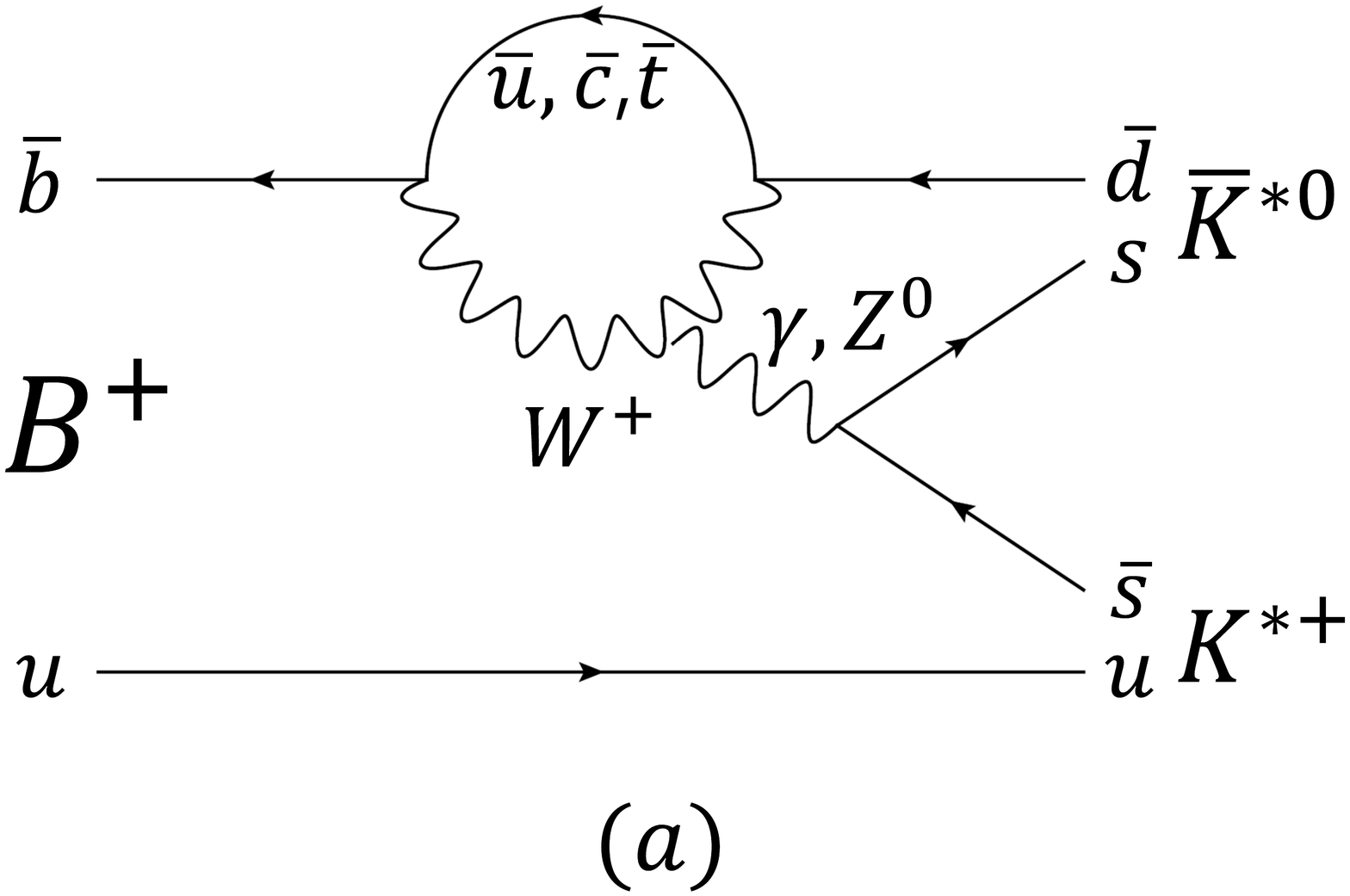}
    \hspace{-0.5cm}
    \includegraphics[width=4.7cm]{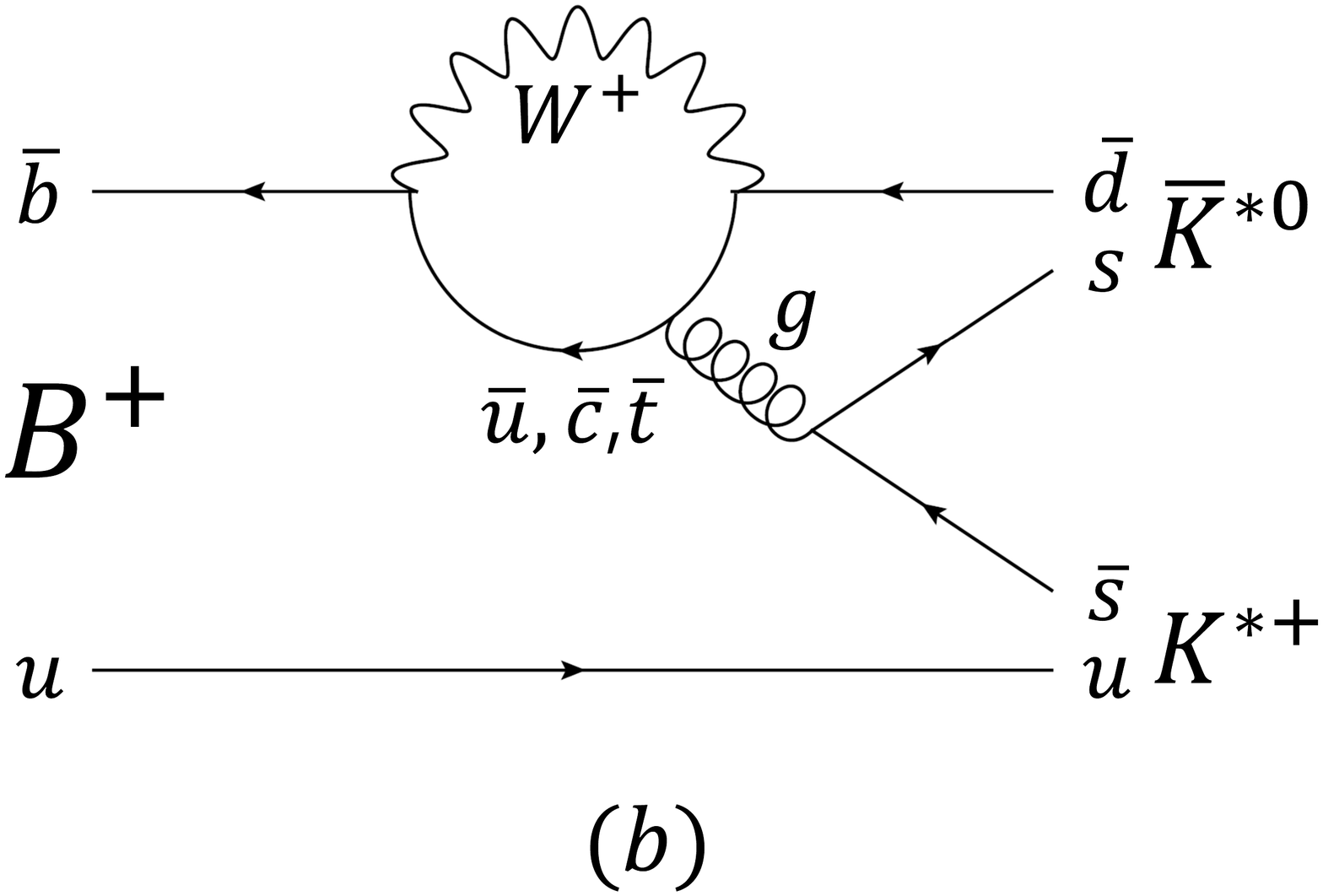}
    \hspace{-0.3cm}
    \vspace{-0.5cm}
    \caption{\footnotesize (a) Electroweak and (b) gluonic $b\rightarrow d$ penguin loop diagrams for $B^+\rightarrow\overline{K}{}^{*0}K^{*+}$. }\label{fig01}
  \end{figure}

The study of charmless $B$ meson decays provides a powerful probe to search for new physics \cite{int00} beyond the standard model.
We search for $B^+\rightarrow\overline{K}{}^{*0}(892)K^{*+}(892)$, a $B\to VV$ decay channel mediated by the $b \to d$ transition for which the so-called polarization puzzle is yet to be solved; here, $V$ denotes a vector meson.
A na\"{i}ve counting rule for light vector mesons predicts the longitudinal-polarization fraction to be $f_L \sim 1-O(m^2_V/m^2_B)$ in such decays \cite{int01}. 
However, in loop-dominated modes such as $B \rightarrow \phi K^{*}$ \cite{int02-2}, the $f_L$ values are found to differ significantly from this prediction. 
In contrast, tree-dominated decays, e.g., $B \rightarrow \rho\rho$ seem to follow the expected pattern \cite{int02-1}.
The polarization puzzle is a prime motivation for measurements in other $B \rightarrow VV$ decays to test predictions of the QCD factorization and  perturbative QCD approach.
The sensitivity to $f_L$ is obtained by considering the decay process in the helicity basis. 
In the $B^+\rightarrow\overline{K}{}^{*0}K^{*+}$ decay, this basis is defined with the two $K^{*}$ rest frames in which the helicity angles $\theta_{K^{*+}}$ and $\theta_{K^{*0}}$ are measured between the daughter momentum ($K^{\pm}$ or $\pi^{\pm}$) of each $K^{*}$ and the direction opposite the $B$ meson.

The $B^+\rightarrow\overline{K}{}^{*0}K^{*+}$ decay proceeds via electroweak and gluonic $b\rightarrow d$ loops, as shown Fig. \ref{fig01}.
The expected branching fractions for $B$ meson decays to $VV$ final states are calculated in several papers \cite{int13, int14, int15, int16, int17, int18, int19}. 
The branching  fraction of $B^+\rightarrow\overline{K}{}^{*0}K^{*+}$  is predicted to be $(0.1-1.1)\times 10^{-6}$ in QCD factorization \cite{int14,int19} and $(0.3-0.9)\times 10^{-6}$ in perturbative QCD \cite{int13,int17}.

The BABAR Collaboration has measured the longitudinal fraction $f_L=0.75^{+0.16}_{-0.26}\pm 0.03$ and the branching fraction $\mathcal{B}=(1.2\pm 0.5 \pm 0.1)\times 10^{-6}$ for $B^+\rightarrow\overline{K}{}^{*0}K^{*+}$ using a data sample of $467\times 10^6$ $B\bar{B}$ pairs \cite{int10}, where the first uncertainty is statistical and the second is systematic. 
It has also obtained the $B^0\rightarrow K^{*0} \overline{K}{}^{*0}$ decay branching fraction $\mathcal{B}=(1.28^{+0.35}_{-0.30} \pm 0.11)\times 10^{-6}$ \cite{int03}. 
On the other hand, Belle reported an upper limit on the branching fraction for $B^0\rightarrow K^{*0} \overline{K}{}^{*0}$ ($B^0\rightarrow K^{*0} K^{*0}$) of $0.81 \times 10^{-6}$ ($0.20 \times 10^{-6}$) \cite{int12}. 
Owing to the smallness of the underlying CKM matrix elements, the $b \to d$ transitions (dominant in $B \to K^{*} K^{*}$ decays) are suppressed compared to $b \to s$ and hence the related channels are not so well measured. 
Therefore, precise measurements based on high statistics are needed to shed more light on the polarization puzzle.

%
%
Our results are based on a data sample containing $772\times10^6$ $B\bar{B}$ pairs, corresponding to an integrated luminosity of $711\, \mathrm{fb}^{-1}$, recorded at the $\Upsilon(4S)$ resonance with the Belle detector \cite{dt01} at the KEKB asymmetric energy $e^+ e^-$ (3.5 on 8.0 GeV) collider \cite{dt02}. The principal detector components used in the study are a silicon vertex detector (SVD), a 50-layer central drift chamber (CDC), an array of aerogel threshold Cherenkov counters (ACC), a barrel-like arrangement of time-of-flight scintillation counters (TOF), and a CsI(Tl) crystal electromagnetic calorimeter (ECL). All these components are located inside a 1.5 T solenoidal magnetic field. Two inner detector configurations were used: a 2.0~cm beampipe and a 3-layer SVD for the first sample of $152 \times 10^{6}$ $B\bar{B}$ pairs, while a 1.5~cm beampipe, a 4-layer SVD and a small-cell CDC to record the remaining $620 \times 10^{6}$ $B\bar{B}$ events  \cite{dt03}. The latter sample has been reprocessed with an improved track reconstruction algorithm, which significantly increased  the signal detection efficiency.

%
%
The $B^+\rightarrow\overline{K}{}^{*0}K^{*+}$ candidate is reconstructed from the subsequent decay channels of  $\overline{K}{}^{*0} \rightarrow K^-\pi^+$ and $K^{*+} \rightarrow K^+ \pi^0$ ($K^0_S \pi^+$), where $K^{*}$ refers to the $K^{*}(892)$ meson \cite{conj}.

%
%
Charged tracks are required to have a transverse momentum greater than $0.1$ $\mathrm{GeV/}c$ and an impact parameter with respect to the interaction point less than $0.3$ $\mathrm{cm}$ in the $r-\phi$ plane and $4.0$ $\mathrm{cm}$ along the $z$ axis. Here, the $z$ axis is the direction opposite the $e^{+}$ beam.
Charged kaons and pions are identified by means of a likelihood ratio $R_{K/\pi}=\mathcal{L}_{K}/(\mathcal{L}_{K}+\mathcal{L}_{\pi})$, where $\mathcal{L}_{K} (\mathcal{L}_{\pi})$ denotes the likelihood for a track being due to a kaon (pion). 
These likelihoods are calculated using specific ionization in the CDC, information from the TOF, and the number of photoelectrons from the ACC.
Kaon identification efficiencies are $98.1\%$ ($99.0\%$) for transversely and $97.2\%$ ($97.5\%$) for longitudinally polarized cases,
and pion identification efficiencies are $97.2\%$ ($98.6\%$) for transversely and $97.3\%$ ($98.9\%$) for longitudinally polarized cases in the $K^{*+} \rightarrow K_S^0 \pi^+$ ($K^{*+}\rightarrow K^+ \pi^0$) channel.
Fake rates for kaons and pions are approximately $0.1\%$ and $0.8\%$, respectively.

%
%
Neutral $\pi^0$ and $K^0_S$ mesons are reconstructed with a pair of photons and charged pions, respectively.
The $\pi^0$ candidates are required to have each daughter photon's energy greater than $0.05$ GeV ($0.10$ GeV) for the barrel (endcap) region of the ECL, a reconstructed invariant mass in the range $0.118$ GeV/$c^2 <m_{\gamma\gamma}<0.150$ GeV/$c^2$, and a $\pi^0$ mass-constrained fit statistic, $\chi^2_{\pi^0}$, smaller than $50$. 
The mass requirement corresponds to $\pm 3\sigma$ around the nominal $\pi^0$ mass~\cite{rfpdg}.
The $K^0_S$ candidates are selected with the following criteria.
The $z$-distance between the two helices at the $\pi^{+}\pi^{-}$ vertex position must be less than $2.5$ cm. 
After this initial selection, the pion momenta are refitted with a common  vertex constraint. The flight length of the $K^0_S$ candidate must lie between $2$ and $20$ cm. The impact parameter with respect to the interaction point must be greater than $0.1$ cm in the $r-\phi$  plane. 
Finally, we require the reconstructed invariant mass to be in the range $0.478$ GeV/$c^2 <m_{\pi\pi}<0.516$ GeV/$c^2$, corresponding to $\pm 5\sigma$ around the nominal $K^0$ mass \cite{rfpdg}.

%
%
The $K^{*}$ candidates are reconstructed by defining the mass range from $0.78$ to $1.00$ $\mathrm{GeV/}c^2$ that corresponds to approximately $\pm 2.1\sigma$ around the nominal $K^*$ mass \cite{rfpdg}. 
In order to reduce the contribution of misreconstructed candidates in the $K^{*+}\rightarrow K^+ \pi^0$ decay, we require the helicity angle of the $K^{*+}$ candidate to satisfy $\mathrm{cos}\,\theta_{K^{*+}}<0.8$. 

%
%
We define two kinematic observables in the form of the energy difference ($\Delta E \equiv E_B-E_{\mathrm{beam}}$) and the beam-energy constrained mass ($M_{\mathrm{bc}}\equiv \frac{1}{c^2} \sqrt{E^2_{\mathrm{beam}}-|\vec{p}_{B}|^2 c^2}$), where $E_\mathrm{beam}$ and $E_B$ ($\vec{p}_B$) are the beam energy and the energy (momentum) of the $B$ meson candidate, respectively, in the $e^+e^-$ center-of-mass (CM) frame.
For the $K^{*+}\rightarrow K^+ \pi^0$ channel, where the $\Delta E$ resolution is poor due to shower leakage in the ECL \cite{int21}, we use the following quantity instead of $M_{\mathrm{bc}}$:
\begin{eqnarray}\label{eqmbc}
&&M^{*}_{\mathrm{bc}}=   \frac{1}{c^2} \Bigg[ E^2_{\mathrm{beam}} \\
&&\left. -\left( \vec{p}_{{}_{\overline{K}{}^{*0}}}c+\frac{\vec{p}_{{}_{K^{*+}}}}{|\vec{p}_{{}_{K^{*+}}}|}\sqrt{(E_{\mathrm{beam}}-E_{\overline{K}{}^{*0}})^2-m^2_{K^{*+}}c^4} \right)^2 \right]^{\frac{1}{2}},\nonumber
\end{eqnarray}
where $m_{K^{*+}}$ is the $K^{*+}$ mass.
We retain $B$ candidates that satisfy $|\Delta E|<0.15$ GeV and $M_{\mathrm{bc}}^{(*)}>5.25$ GeV$/c^2$.

%
%
The dominant background arises from the $e^+ e^-\rightarrow q\bar{q}~(q=u,d,s,c)$ continuum process. To suppress these events, a neural network \cite{int26} is employed by combining the following four quantities: a Fisher discriminant  formed from 16 modified Fox-Wolfram moments \cite{int27}, the cosine of the angle between the momentum of signal $B$ candidate and the $z$ axis in the CM frame, the separation along the $z$ axis between the vertex of the signal $B$ and that of the recoil $B$, and the recoil $B$'s flavor-tagging information \cite{int28}. 
To reconstruct the decay vertex of the recoil $B$, the tracks not associated with the signal $B$ are used.
The training and optimization of the neural network are accomplished with signal and continuum Monte Carlo (MC) simulated events. 
The signal MC sample is generated with the EvtGen program \cite{int29}, taking final-state radiation effects into account via PHOTOS \cite{int_photos}.
The neural network output ($C_{NB}$) ranges from $-1$ to $+1$; an event near $+1$ ($-1$) is more signal (continuum)-like.
We require $C_{NB}>-0.5$ to reduce substantially the amount of continuum background.
This requirement preserves approximately $94.7\%$ ($94.5\%$) of the signal while suppressing $75.6\%$ ($71.2\%$) of the continuum background in $K^{*+} \rightarrow K^+\pi^0$ ($K^{*+} \rightarrow K_S^0 \pi^+$). 
As the remainder of the $C_{NB}$ distribution has a sharp peak near unity, we use a  transformed quantity to enable its modeling with an analytic shape:
\begin{eqnarray}
C^{\prime}_{NB} = \mathrm{log} \left( \frac{C_{NB}-C_{NB}^{\rm{min}}}{C_{NB}^{\rm{max}}-C_{NB}} \right) \label{nb01},
\end{eqnarray}
where $C_{NB}^{\rm{min}}=-0.5$ and $C_{NB}^{\rm{max}}=0.997$ ($0.995$) in $K^{*+} \rightarrow K^+\pi^0$ ($K^{*+} \rightarrow K_S^0 \pi^+ $).

%
%
\begin{table*}[t]
\caption{\footnotesize List of PDFs used to model the $\Delta E$, $M_{\mathrm{bc}}^{(*)}$, $m_{K\pi}$, $m_{K^0_S\pi(K^+\pi^0)}$, $\mathrm{cos\theta}_{K\pi}$, $\mathrm{cos\theta}_{K_S\pi(K^+\pi^0)}$ and $C_{NB}^{\prime}$ distributions for the various event categories. G, AG, CB, ARG, (r)BW, $\textrm{P}_i$, LASS, Hist and Erf stand for Gaussian, asymmetric Gaussian, Crystal Ball \cite{int31}, ARGUS function \cite{int32}, (relativistic) Breit-Wigner function, $i$-th order Chebyshev polynomial, LASS parameterization for the $K_0^{*}(1430)$ line shape, histogram and error function, respectively. Two different PDFs are used to model $\mathrm{cos\theta}_{K\pi}$ on the two samples of $m_{K\pi}<0.83$ GeV/$c^2$ and $m_{K\pi}>0.83$ GeV/$c^2$. }\label{sum_table_pdf}
\centering
\begin{tabular*}{\textwidth}{@{\extracolsep{\fill}}c|cccccccc}
  \hline
  \hline
  Final state & Event category &  $\Delta E$ & $M_{bc}^{(*)}$ & $m_{K\pi}$ & $m_{K^0_S\pi(K^+\pi^0)}$ & $\mathrm{cos\theta}_{K\pi}$ & $\mathrm{cos\theta}_{K^0_S\pi(K^+\pi^0)}$ & $C_{NB}^{\prime}$ \\
  \hline
  & Signal (RC)   & 2G  &  G    &  rBW  &  rBW  &  Hist / Hist  &  Hist  &  2AG \\
  & Signal (SCF) & Hist &  Hist &  Hist  &  Hist  &  Hist / Hist  &  Hist  &  AG   \\
  & Continuum $q\bar{q}$ & $\textrm{P}_1$     &  ARG    &  rBW+$\textrm{P}_1$  &  rBW+$\textrm{P}_1$  &  $\textrm{P}_6$$\times$Erf / $\textrm{P}_4$$\times$Erf  &  $\textrm{P}_5$$\times$Erf  &  2G \\
  & Charm $B\bar{B}$     &  $\textrm{P}_1$    &  ARG    &  $\textrm{P}_1$         &  $\textrm{P}_2$         &  $\textrm{P}_4$ / $\textrm{P}_4$$\times$Erf & $\textrm{P}_5$        &  AG \\
  $K^- \pi^+ K^0_S \pi^+$ & Charmless $B\bar{B}$                      & G+$\textrm{P}_2$ & G+ARG & BW+$\textrm{P}_1$    &  BW+$\textrm{P}_1$   & Hist / Hist                                     &  Hist                                     &  AG \\
  & $B^+ \rightarrow (\overline{K\pi})_0^{*0} K^{*+}$    &  2G & 2G  & LASS & rBW & Hist / Hist & Hist & AG \\
  & $B^+ \rightarrow \overline{K}{}_2^{*0} K^{*+}$         &  2G & 2G  & BW    & rBW & Hist / Hist & Hist & AG \\
  & $B^+ \rightarrow \overline{K}{}^{*0} (K\pi)_0^{*+}$ &  G+$\textrm{P}_2$  & G & rBW  & LASS  & Hist / Hist & Hist & AG \\
  & $B^+ \rightarrow \overline{K}{}^{*0} K_2^{*+}$         &  G+$\textrm{P}_2$  & G & rBW  & BW     & Hist / Hist & Hist & AG \\
  & $B^+ \rightarrow$ four-body      &  G+$\textrm{P}_2$  & G+$\textrm{P}_2$  & $\textrm{P}_1$ & $\textrm{P}_1$ & Hist / Hist & Hist & AG \\
  \hline
  & Signal (RC)   & CB+G  &  CB  &  rBW  &  rBW  &  Hist / Hist  &  Hist  &  2AG \\
  & Signal (SCF) & Hist     &  Hist &  Hist  &  Hist  &  Hist / Hist  &  Hist  &  AG   \\
  & Continuum $q\bar{q}$ & $\textrm{P}_1$   &  ARG    &  rBW+$\textrm{P}_1$  &  rBW+$\textrm{P}_1$  &  $\textrm{P}_6$$\times$Erf / $\textrm{P}_5$$\times$Erf &  $\textrm{P}_6$ &  2AG \\
  & Charm $B\bar{B}$     & $\textrm{P}_2$   &  ARG    &  $\textrm{P}_1$         &  $\textrm{P}_1$         &  $\textrm{P}_4$ / $\textrm{P}_4$$\times$Erf & $\textrm{P}_4$  &  AG \\
  $K^- \pi^+ K^+ \pi^0$ & Charmless $B\bar{B}$                      & $\textrm{P}_4$   & $\textrm{P}_4$ & BW+$\textrm{P}_1$    &  BW+$\textrm{P}_1$   & Hist / Hist  &  Hist  &  AG \\
  & $B^+ \rightarrow (\overline{K\pi})_0^{*0} K^{*+}$  &  CB+$\textrm{P}_2$ & 2G  & LASS & rBW & Hist / Hist & Hist & 2AG \\
  & $B^+ \rightarrow \overline{K}{}_2^{*0} K^{*+}$         &  CB+$\textrm{P}_2$ & 2G  & BW    & rBW & Hist / Hist & Hist & 2AG \\
  & $B^+ \rightarrow \overline{K}{}^{*0} (K\pi)_0^{*+}$  &  CB+$\textrm{P}_2$ & 2G  & rBW  & LASS  & Hist / Hist & Hist & 2AG \\
  & $B^+ \rightarrow \overline{K}{}^{*0} K_2^{*+}$         &  CB+$\textrm{P}_2$ & 2G  & rBW  & BW     & Hist / Hist & Hist & 2AG \\
  & $B^+ \rightarrow $ four-body   &  G+$\textrm{P}_2$  & G+$\textrm{P}_2$  & $\textrm{P}_1$ & $\textrm{P}_1$ & Hist / Hist & Hist & AG \\
  \hline
  \hline
\end{tabular*}
\end{table*}

%
%
After all selection criteria are applied to the signal MC sample, the average number of signal candidates per event is $1.16$ ($1.13$) for longitudinally (transversely) polarized decays in $K^{*+} \rightarrow K^+\pi^0$ and $1.10$ ($1.06$) in $K^{*+} \rightarrow K_S^0 \pi^+$. 
We choose the candidate having the smallest $\chi_{\pi^0}^2+\chi_{B}^2$ ($\chi_{K_S^0}^2+\chi_{B}^2$) value in $K^{*+} \rightarrow K^+\pi^0$ ($K^{*+} \rightarrow K_S^0 \pi^+$), where the $B$ vertex is obtained by charged tracks except for those from $K^0_S$ and $\chi^2_{B}$ ($\chi^2_{K_S^0}$) is the $B$ ($K_S^0$) vertex-fit statistic.
We refer to the right-combination (RC) as the correctly reconstructed $B$ meson decays and the self-crossfeed (SCF) as the misreconstructed signal component.
MC simulations show that the SCF fraction is $15.5\%$ ($10.2\%$) for the longitudinally (transversely) polarized case in $K^{*+} \rightarrow K^+\pi^0$ and $7.7\%$ ($3.5\%$) for the longitudinally (transversely) polarized $K^{*+} \rightarrow K_S^0 \pi^+$ decay.

%
%
The charm $B\bar{B}$ background originating from the $b\rightarrow c$ transition remains after all event selection criteria are applied. 
In the MC sample, we find no peaking structure in $\Delta E$, $M_{\mathrm{bc}}^{(*)}$, and the invariant masses formed by combining two or three final-state particles.
We also do not observe any specific charm decay mode in this sample.
The other possible backgrounds are largely due to $b\rightarrow u,d,s$ transitions from charmless $B$ decays. 
This background has no peaking structure in the signal enhanced region of $|\Delta E|<0.05$ GeV,
while a peaking structure originated from $B^{+}\rightarrow \rho^{0} K^{*+}$ and $B^{+}\rightarrow \pi\pi K^{*+}$ with $K^{*+} \rightarrow K_S^0 \pi^+$  is seen at $\Delta E \sim 0.07$ GeV.
Other backgrounds involving higher $K^{*}$ states such as $K^{*} K_{2}^{*}(1430)$ and $K^{*} K_{0}^{*}(1430)$, $K\pi K^{*}$ decays, and the nonresonant four-body $K \pi K^0_S \pi$ ($K \pi K \pi^0$) decays also contribute. 
The  $K^{*} K_{2}^{*}(1430)$ decays are simulated based on the theoretical expectations \cite{int30} for branching fractions and polarizations.
The contributions of $K^{*} K_{0}^{*}(1430)$ decays are estimated on both $K^{*}$ mass sidebands, where the $K_S^0\pi$ $(K\pi^0)$ mass sideband is $0.78$ GeV/$c^2<m_{K\pi}<1.00$ GeV/$c^2$ and $1.00$ GeV/$c^2<m_{K_S^0\pi(K\pi^0)}<1.52$ GeV/$c^2$ and the $K\pi$ mass sideband is $1.00$ GeV/$c^2<m_{K\pi}<1.52$ GeV/$c^2$ and $0.78$ GeV/$c^2<m_{K_S^0\pi(K\pi^0)}<1.00$ GeV/$c^2$.
The $B^+ \rightarrow \phi K^{*+}$ background arising from pion-to-kaon misidentification is suppressed by rejecting events with an invariant mass of the $K^{+}K^{-}$ pair between $1006.5$ and $1032.5$ MeV/$c^2$.

%
%
The $B \rightarrow VV$ decay rate does not depend strongly on the azimuthal angle, $\phi$, between the two decay planes of the vector mesons. 
Therefore, it can be integrated out to obtain the differential decay rate \cite{hel01}
\begin{eqnarray}\label{eq001}
\frac{1}{\Gamma} \frac{d^2 \Gamma}{d \mathrm{cos}\theta_{\overline{K}{}^{*0}} d \mathrm{cos}\theta_{ K^{*+} } } & = &
\frac{9}{16}(1-f_L) \mathrm{sin}^2\theta_{\overline{K}{}^{*0}} \mathrm{sin}^2\theta_{K^{*+}}  \nonumber\\
&&+ \frac{9}{4} f_L \mathrm{cos}^2\theta_{\overline{K}{}^{*0}}  \mathrm{cos}^2\theta_{K^{*+}} \,\,\,.
\end{eqnarray}

%
%
We obtain the branching fraction $\mathcal{B}$ and the longitudinal polarization fraction $f_L$ using a simultaneous fit to the $K^{*+} \rightarrow K_S^0 \pi^+$ and $K^{*+}\rightarrow K^+ \pi^0$ decay channels.
This is an unbinned extended maximum likelihood (ML) fit to the distributions of $\Delta E$ and $M_{\mathrm{bc}}^{(*)}$, the invariant mass and the cosine of the helicity angle of the two $K^{*}$ candidates, and $C_{NB}^{\prime}$. 
The extended ML function for each decay channel is
\begin{eqnarray}
\mathcal{L}=\frac{1}{N!}\textrm{exp}\left( -\sum_j n_j  \right) \times \prod_{i=1}^N \left[ \sum_j n_j \mathcal{P}_j (\vec{x}_i ; \vec{\alpha}_j )   \right],
\end{eqnarray}
where $\mathcal{P}_j (\vec{x}_i ; \vec{\alpha}_j )$ is the product of uncorrelated one-dimensional (1D) probability density functions (PDFs)  for event category $j$, calculated for the seven measured observables $\vec{x}_i$ of the $i$-th event, 
$n_j$ is the yield for this event category, and $N$ is the total number of events. 
The parameters $\vec{\alpha}_j$ describe the expected distributions of the measured observables for event category $j$, and are extracted from MC simulations and the ($K^{*}$ mass, $M_{\mathrm{bc}}^{(*)}$) sideband data.
For the simultaneous fit, the total likelihood is obtained by multiplying the likelihoods for  the $K^{*+} \rightarrow K_S^0 \pi^+$ and $K^{*+}\rightarrow K^+ \pi^0$ decay channels  (indexed by $k$).
With an assumption of equal production of $B^+ B^-$ and $B^0 \bar{B}^0$ pairs at the $\Upsilon (4S)$ resonance,
the signal yield of channel  $k$ is given by $n_{\rm{sig},\it{k}}=\mathcal{B} \times \left[ f_L \epsilon^L_{\rm{rec},\it{k}} + (1-f_L)  \epsilon^T_{\rm{rec},\it{k}} \right] \times  \Pi \mathcal{B}_k \times  N_{B\bar{B}}$, 
where $N_{B\bar{B}}$ is the number of $B\bar{B}$ pairs, $n_{\textrm{sig}}$ is the number of signal events, and $\Pi \mathcal{B}_k$ is the product of the sub-branching fractions.
The detection efficiency for the longitudinally (transversely) polarized mode, $\epsilon^{L(T)}_{\textrm{rec}}$, is equal to  $11.58 \pm 0.02 \%$ ($14.41 \pm 0.02 \%$) and $12.35 \pm 0.02 \%$ ($17.29 \pm 0.02 \%$) for  the $K^{*+} \rightarrow K_S^0 \pi^+$ and $K^{*+}\rightarrow K^+ \pi^0$ channels, respectively. 
These are determined primarily from the signal MC sample and then corrected for a modest difference of kaon-identification efficiency between data and simulations, given by $r_{K/\pi} \equiv \varepsilon^{\textrm{data}}_{K/\pi} / \varepsilon^{\textrm{MC}}_{K/\pi}$, where $\varepsilon^{\textrm{data}}_{K/\pi}$ ($ \varepsilon^{\textrm{MC}}_{K/\pi}$) is the efficiency of the $R_{K/\pi}$ requirement in data (simulations).
The $r_{K/\pi}$ value per charged pion (kaon) track is $0.96$ ($1.00$), resulting in a total efficiency of $0.92$ ($0.96$) for  $K^{*+} \rightarrow K_S^0 \pi^+$ ($K^{*+}\rightarrow K^+ \pi^0$).
Though mild linear correlations of up to $15\%$ exist in the signal, such as between ($\Delta E$, $M_{\mathrm{bc}}$), their contributions to the fit bias (described later) due to our use of uncorrelated 1D PDFs are negligible.

Table \ref{sum_table_pdf} lists the PDF shapes used to model $\Delta E$, $M_{\mathrm{bc}}^{(*)}$, $m_{K\pi}$, $m_{K^0_S\pi(K\pi^0)}$, $\mathrm{cos\theta}_{K\pi}$, $\mathrm{cos\theta}_{K_S\pi(K\pi^0)}$ and $C_{NB}^{\prime}$ for different event categories.
We fix the parameters of the RC signal PDF shapes to the MC values. We correct the parameters of the RC signal $\Delta E$, $M_{\mathrm{bc}}^{(*)}$ and $C_{NB}^{\prime}$ PDFs to account for modest data-MC 
differences; the correction factors are obtained from  a high-statistics control sample of $B^{+} \rightarrow J/\psi (\mu^+\mu^-) K^{*+}$. 
The same calibration factors are also applied to the higher-$K^{*}$ and nonresonant backgrounds.

The continuum background PDF parameters that are allowed to vary are the slope of $\Delta E$, the shape of $M_{\mathrm{bc}}^{(*)}$, the fraction of the relativistic Breit-Wigner function, the polynomial coefficients of  the $K^{*}$ masses, and the mean and two widths for the core asymmetric Gaussian function of $C_{NB}^{\prime}$. 
All other PDF parameters are fixed and determined from MC samples.
We use an error function to describe the falling reconstruction efficiency due to low-momentum tracks in the continuum as well as the $B\bar{B}$ helicity angle distributions.
We use the simultaneous fit with two different $\mathrm{cos\theta}_{K\pi}$ PDFs, corresponding to the two samples of $m_{K\pi}<0.83$ GeV/$c^2$ and $m_{K\pi}>0.83$ GeV/$c^2$, to treat the correlation between $m_{K\pi}$ and $\mathrm{cos\theta}_{K\pi}$ that originates from the $B \rightarrow \phi K^{*}$ veto.

\begin{figure*}[t]
   \centering
   \includegraphics[width=\textwidth]{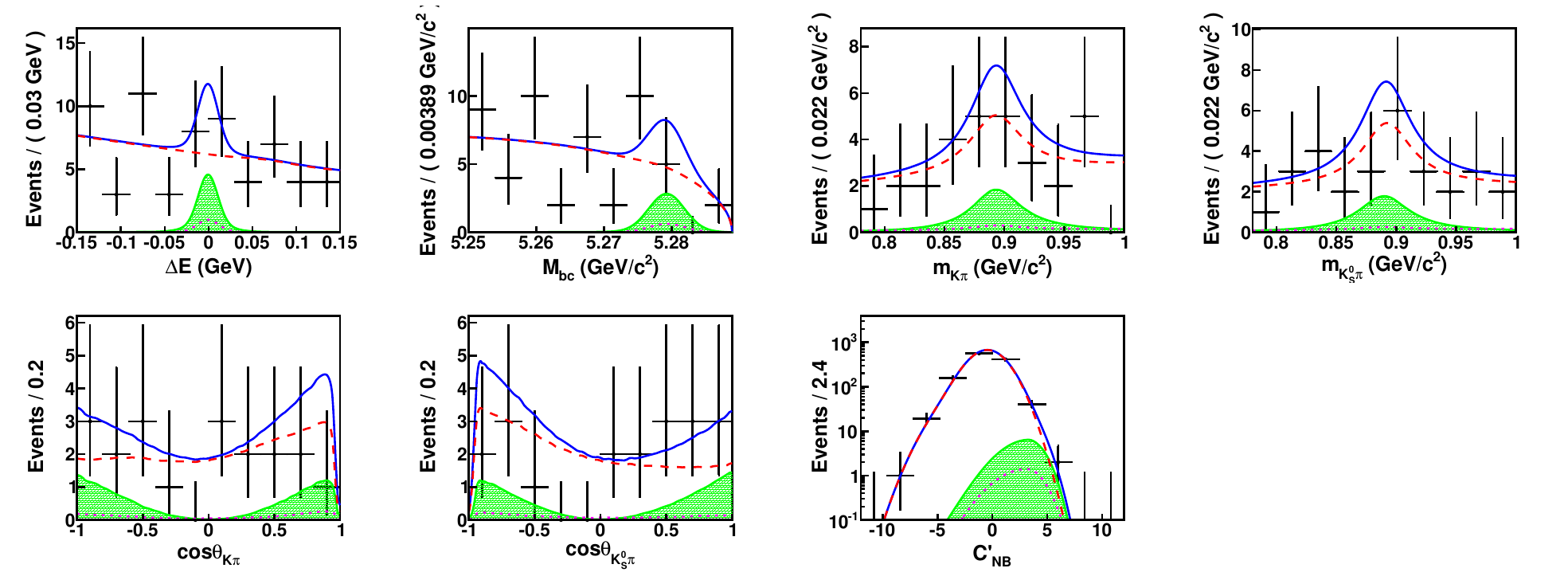}
   \caption{\footnotesize (color online). Projections for $B^+\rightarrow\overline{K}{}^{*0}(\rightarrow K^{-}\pi^{+}) K^{*+}(\rightarrow K^{0}_{S}\pi^{+})$ of the multidimensional fit onto $\Delta E$, $M_{bc}$, $\overline{K}{}^{*0}$ mass, $K^{*+}$ mass, cosine of $\overline{K}{}^{*0}$  helicity angle, cosine of $K^{*+}$  helicity angle, and $C_{NB}^{\prime}$ for events selected in a signal enhanced region with the plotted variable excluded. Points with error bars are the data, the solid curves represent the full fit function, the hatched regions are the signal, the dashed curves show the combined continuum and $B\bar{B}$  backgrounds, and the dotted curves are the higher $K^{*}$ and nonresonant backgrounds.}\label{fig02}
\end{figure*}

The yields for all event categories except for the relative amount of SCF to RC signal, the charmless $B\bar{B}$, higher $K^{*}$ and nonresonant background components are allowed to vary in the fit. 
We fix the yields of charmless $B\bar{B}$ backgrounds based on a high-statistics MC sample, which includes possible charmless rare $B$ decays.
In order to validate our fitting procedure, we perform the fit to ensembles of 500 pseudoexperiments using the extracted fitted yields from data and events of all components that are arbitrarily chosen from the simulated MC samples. 
We obtain and correct for fit biases of $1.8\%$ and $8.2\%$ for $\mathcal{B}$ and $f_L$, respectively, and assign $50\%$ of each bias as its systematic uncertainty.

The $K_{0}^{*}(1430)$ resonance, together with an effective-range nonresonant component, are modeled with the LASS function \cite{int33}, whose parameters are taken from Ref. \cite{int34}.
Yields of $(\overline{K\pi})_0^{*0}K^{*+}$, $\overline{K}{}^{*0}(K\pi)_0^{*+}$ and four-body decay backgrounds are measured by a simultaneous fit to the sidebands of the two $K^{*}$ masses.
To combine the results of the two $K^{*+}$ decay channels, both fits share the branching fraction parameters of $(\overline{K\pi})_0^{*0}K^{*+}$, $\overline{K}{}^{*0}(K\pi)_0^{*+}$ and four-body decay backgrounds for $K^{*+} \rightarrow K_S^0 \pi^+$  and $K^{*+} \rightarrow K^+\pi^0$ in the simultaneous fit.
In the fit, these background yields in the $K^{*}$ mass signal region from $0.78$ to $1.00$ GeV/$c^2$ are estimated from the $K^{*}$ mass PDFs on the two $K^{*}$ mass sidebands. 

We obtain the biases of the $(K\pi)_0^{*}K^{*}$ and four-body decay yields by applying the fit to ensembles of 500 pseudoexperiments using the extracted fitted yields from the $K^{*}$ mass sidebands. 
Fit biases for the yields of $(\overline{K\pi})_0^{*0}K^{*+}$, $\overline{K}{}^{*0}(K\pi)_0^{*+}$ and four-body decays are, respectively, $3.0$ ($2.6$), $2.0$ ($2.0$) and $0.8$ ($0.4$)  in the $K^{*+} \rightarrow K_S^0 \pi^+$ ($K^{*+} \rightarrow K^+\pi^0$) sample.
We correct for the fit biases and assign $50\%$ of each to the systematic uncertainties.
The measured yields in the $K^{*}$ mass sidebands are extrapolated to the $K^{*}$ mass signal region using the $K^{*}$ mass PDFs. 
We obtain the background yields $N_{(\overline{K\pi})_0^{*0}K^{*+}}=1.9^{+2.9}_{-2.8}$ ($1.6^{+2.5}_{-2.4}$), $N_{\overline{K}{}^{*0}(K\pi)_0^{*+}}=3.3^{+2.7}_{-2.3}$ ($3.2\pm 1.9$), and $N_{\textrm{4body}}=2.5\pm 3.0$ ($1.2\pm 1.4$) in the $K^{*+} \rightarrow K_S^0 \pi^+$ ($K^{*+} \rightarrow K^+\pi^0$) samples, where errors are a quadratic sum of the statistical and systematic uncertainties.

\begin{figure*}[t]
   \centering
   \includegraphics[width=\textwidth]{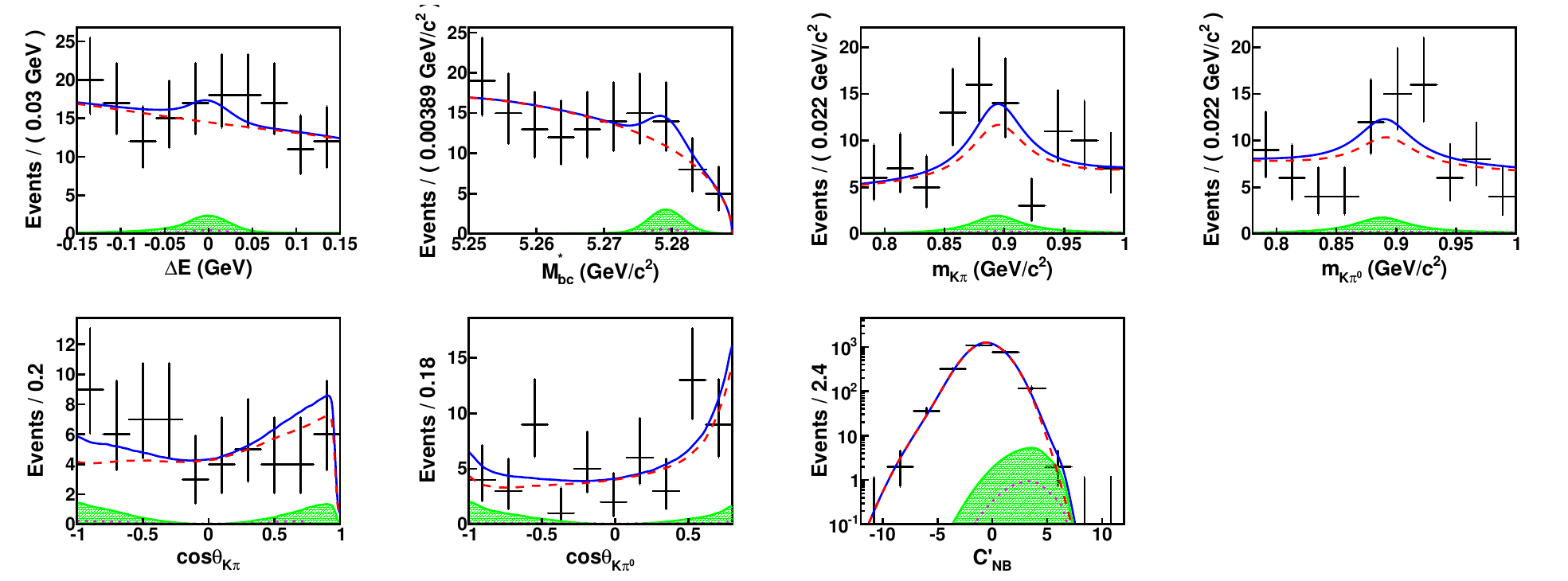}
   \caption{\footnotesize (color online). Projections for $B^+\rightarrow\overline{K}{}^{*0}(\rightarrow K^{-}\pi^{+}) K^{*+}(\rightarrow K^{+}\pi^{0})$ of the multidimensional fit onto $\Delta E$, $M^{*}_{\mathrm{bc}}$, $\overline{K}{}^{*0}$ mass, $K^{*+}$ mass, cosine of $\overline{K}{}^{*0}$  helicity angle, cosine of $K^{*+}$  helicity angle, and $C_{NB}^{\prime}$. The same projection criteria and legend as Fig. \ref{fig02} are used.}\label{fig03}
\end{figure*}

\begin{table}[bp]
\caption{\footnotesize Summary of results for the fitted yields, average efficiencies $\epsilon_{\textrm{rec}}$ for the fitted $f_L$, sub-branching fractions $\prod \mathcal{B}$, longitudinal polarization fraction $f_L$, branching fraction $\mathcal{B}(B^+\rightarrow\overline{K}{}^{*0} K^{*+})$, signal significance $S$, and $\mathcal{B}$ upper limit at $90\%$ CL. The first error is statistical and the systematic error is quoted last, if given.}\label{sum_datafit}
\centering
\begin{tabular*}{\columnwidth}{@{\extracolsep{\fill}}lcc}
  \hline
  \hline
  Final state & $K^- \pi^+ K^0_S \pi^+$ &   $K^- \pi^+ K^+ \pi^0$ \\
  \hline	
  Yields (events): & & \\
  Total                                      & 23338 & 50212 \\
  Signal                                    & $15.8^{+7.2}_{-6.1}$ & $16.7^{+7.6}_{-6.5}$ \\
  $q\bar{q}$          & $22982^{+213}_{-212}$  & $49733^{+276}_{-278}$ \\
  Charm $B\bar{B}$          & $265^{+151}_{-149}$  & $290^{+168}_{-162}$ \\
  Charmless $B\bar{B}$ (fixed)     & $78$ & $166$  \\
  $(\overline{K\pi})_{0}^{*0} K^{*+}$ (fixed) & $1.9$ & $1.6$  \\
  $\overline{K}{}^{*0} (K\pi)_{0}^{*+}$ (fixed) & $3.3$ & $3.2$  \\
  $\overline{K}{}_{2}^{*0} K^{*+}$ (fixed)         & $0.45$ & $0.30$  \\
  $\overline{K}{}^{*0} K_{2}^{*+}$ (fixed)         & $0.10$ & $0.06$  \\
  \vspace{0.15cm}
  four-body decay                                             & $2.5$ & $1.2$  \\   
  Efficiencies:  & & \\
  $\epsilon_{\textrm{rec}}(\%)$ & $11.58 \pm 0.02$  & $12.35 \pm 0.02$ \\
  \vspace{0.15cm}
  $\prod \mathcal{B}_i (\%)$ & $15.37$ & $21.96$ \\ 
  \hline
  Results:  & & \\
  $f_L$ &  \multicolumn{2}{c}{$1.06 \pm 0.30 \pm 0.14$} \\
  $\mathcal{B}(\times 10^{-6})$ &  \multicolumn{2}{c}{$0.77^{+0.35}_{-0.30} \pm 0.12$} \\
  $S(\sigma)$ & \multicolumn{2}{c}{$2.7$} \\
  $\mathcal{B}(\times 10^{-6})$ upper limit ($90\%$ C.L.) & \multicolumn{2}{c}{$1.31$} \\
  \hline
  \hline
\end{tabular*}
\end{table}

The total event sample for $B^+\rightarrow\overline{K}{}^{*0} K^{*+}$ consists of $23338$ and $50212$ events with  $K^{*+} \rightarrow K_S^0 \pi^+$ and $K^{*+}\rightarrow K^+ \pi^0$, respectively.
The result of the ML fit is summarized in Table \ref{sum_datafit}.
We take the sub-branching fractions $\mathcal{B}(\overline{K}{}^{*0} \rightarrow K^{-}\pi^{+})=2/3$, $\mathcal{B}(K^{*+} \rightarrow K^{0}\pi^{+})=2/3$, $\mathcal{B}(K^{*+} \rightarrow K^{+}\pi^{0})=1/3$ and $\mathcal{B}(K^{0} \rightarrow K_S^0  \rightarrow \pi^{+}\pi^{-})=0.5\times (69.20\pm 0.05)\%$ \cite{rfpdg}. 
The signal significance $S$ is defined as $\sqrt{-2 \textrm{log} (\mathcal{L}_{\textrm{max}}/\mathcal{L}_{0})}$, where $\mathcal{L}_{\textrm{max}}$ ($\mathcal{L}_{0}$) is the likelihood value when the signal yield is set to its nominal value (zero).
The systematic uncertainty (discussed below) is included in this significance calculation by convolving the statistical likelihood with an asymmetric Gaussian distribution whose width equals the total systematic error. 
The total significance of the signal yield is $2.7$ standard deviations ($\sigma$). 
The upper limit (UL) on the branching fraction is calculated at $90\%$ confidence-level (CL) by using the formula $\int^{\mathcal{B}_{\textrm{UL}}}_{0} \mathcal{L}(\mathcal{B})d\mathcal{B}/\int^{\infty}_{0} \mathcal{L}(\mathcal{B})d\mathcal{B}=0.9$.
The result is $\mathcal{B}_{\mathrm{UL}} = 1.31\times 10^{-6}$.

Figures \ref{fig02} and \ref{fig03} show the projections of the two fits onto $\Delta E$, $M^{(*)}_{\mathrm{bc}}$, $K^{*}$ masses, cosines of the helicity angle and $C_{NB}^{\prime}$ for $K^{*+} \rightarrow K_S^0 \pi^+$ and $K^{*+}\rightarrow K^+ \pi^0$. 
The candidates and PDFs in each figure are projected in the signal-enhanced region: $|\Delta E|<0.05$ GeV, $M_{\mathrm{bc}}^{(*)}>5.27$ GeV$/c^2$, $0.83$ GeV$/c^2<m_{K^{*}}<0.95 $ GeV$/c^2$ and $C_{NB}^{\prime}>3$.

\begin{table}[bp]
\caption{\footnotesize Summary of systematic uncertainties (\%) on the branching fraction and longitudinal polarization fraction. }\label{sum_syst}
\centering
\begin{tabular*}{\columnwidth}{@{\extracolsep{\fill}}lcc}
  \hline
  \hline
    & $\mathcal{B}$ & $f_L$ \\
  \hline	
  Fit bias & $4.72$ & $6.81$ \\
  PDF modeling & $5.40$ & $5.32$ \\
  Histogram PDF & $1.07$ & $1.11$ \\
  Calibration factors & ${}^{+5.32}_{-3.82}$ & - \\
  Track reconstruction & $2.10$ & - \\
  PID efficiency& $3.90$ & - \\
  $C_{NB}$ efficiency & $1.08$ & - \\
  \vspace{0.15cm}
  $K^0_S$ reconstruction & $0.73$ & - \\
  $\pi^0$ reconstruction & $4.09$ & - \\
  Fractions of misreconstructed events & ${}^{+3.32}_{-1.48}$ & ${}^{+1.92}_{-2.80}$ \\
  Nonresonant \& higher $K^{*}$ background &${}^{+9.54}_{-9.73}$ & ${}^{+3.76}_{-4.10}$ \\
  Limited MC statistics & $0.31$ & - \\
  Charmless $B\bar{B}$ background & ${}^{+2.13}_{-0.67}$ & - \\
  Number of $B\bar{B}$ events& $1.37$ & - \\
  Interference with $(K\pi)_{0}^{*}$ & $5.80$ & $9.69$ \\
  \hline
  Total & ${}^{+16.2}_{-15.4}$ & ${}^{+13.7}_{-13.9}$ \\
  \hline
  \hline
\end{tabular*}
\end{table}

%
%
One of the sources for fit bias is its inaccurate estimation (based on the ensemble test) due to the limited size of the $q\bar{q}$ MC samples.
%
%
%
%
The MC samples are generated under a partially correlated $q\bar{q}$ PDF. 
We estimate an additional fit bias from the results of a comparison between ensemble tests using limited and correlated MC samples.
We calculate the total fit bias uncertainty as the quadratic sum with this additional fit bias.
The uncertainties due to the fixed yields for the higher $K^{*}$ and nonresonant backgrounds are estimated by varying the corresponding yields by their errors. 
The uncertainties due to the fixed fractions of misreconstructed events are calculated by varying them by $\pm50\%$.
The charmless $B\bar{B}$ background yield is also varied by a conservative $\pm 50\%$ to cover any mismodeling of these processes in the MC sample used to estimate the yield.
The change in the signal yield is taken as the systematic uncertainty.

We estimate the effect of possible interference between the $K^{*}$  and spin-0 final states [nonresonant and $K_0^{*}(1430)$] by including interference terms with variable phases in the relativistic Breit-Wigner function of the spin-0 final-state mass.
In this estimation, we assume the $K^{*}$ helicity angle distributions for $f_L=0$ and $f_L=1$ in the $K\pi K^{*}$ decay to be the same as those of our signal decay.
We vary the amplitude and phase of the interference term and the fractions of $f_L=0$ and $f_L=1$ components of $K\pi K^{*}$ from $0$ to $1$.
We assign the resulting shifts as the systematic uncertainties after refitting with this modified function. 

We vary the bin height for all histogram PDFs by its statistical error and repeat the fit. 
Deviations from the nominal fit are added in quadrature to determine the uncertainty due to the histogram PDFs.
The PDF modeling uncertainty is obtained by varying the fixed shape parameters by their errors. 
We assign an uncertainty on the absolute scale of the reconstruction efficiency due to the limited signal MC statistics.
The uncertainty due to calibration factors to correct for the difference between data and simulations is obtained by varying those factors by their errors. 
We assign an uncertainty due to the different continuum suppression efficiencies at $C_{NB}=-0.5$ in data and MC by using the $B^{+} \rightarrow J/\psi (\mu^+\mu^-) K^{*+}$ control sample. 
We also include reconstruction efficiency uncertainties for charged tracks (0.35\% per track) by using partially reconstructed $D^{*+}\rightarrow D^0 (K_S^0 \pi^+ \pi^-) \pi^+$, particle identification (PID) uncertainties by using the $D^{*+}\rightarrow D^0 (K^- \pi^+) \pi^+$ control sample, and the uncertainty on the number of $B\bar{B}$ pairs.
The systematic uncertainty due to the $\pi^0$ reconstruction is obtained by comparing data-MC differences of the yield ratio between $\eta \rightarrow \pi^0 \pi^0 \pi^0$ and $\eta \rightarrow \pi^+ \pi^- \pi^0$.
The systematic uncertainties  on the branching fraction and longitudinal polarization are listed in Table~\ref{sum_syst}.

%
%
In summary, we have searched for the charmless hadronic decay $B^+\rightarrow\overline{K}{}^{*0}K^{*+}$ using the full $B\bar{B}$ pair sample collected with Belle.
We find a $2.7\sigma$ excess of signal with a branching fraction $\mathcal{B}=(0.77^{+0.35}_{-0.30} \pm 0.12)\times 10^{-6}$ and a longitudinal polarization fraction $f_L=1.06 \pm 0.30 \pm 0.14$. 
We obtain a branching fraction upper limit of $1.31\times 10^{-6}$ at $90\%$ CL.

%
%
We thank the KEKB group for excellent operation of the
accelerator; the KEK cryogenics group for efficient solenoid
operations; and the KEK computer group, the NII, and 
PNNL/EMSL for valuable computing and SINET4 network support.  
We acknowledge support from MEXT, JSPS and Nagoya's TLPRC (Japan);
ARC and DIISR (Australia); FWF (Austria); NSFC (China); MSMT (Czechia);
CZF, DFG, and VS (Germany); DST (India); INFN (Italy); 
MOE, MSIP, NRF, GSDC of KISTI, and BK21Plus (Korea);
MNiSW and NCN (Poland); MES and RFAAE (Russia); ARRS (Slovenia);
IKERBASQUE and UPV/EHU (Spain); 
SNSF (Switzerland); NSC and MOE (Taiwan); and DOE and NSF (USA).

\end{document}

%% file: author.tex
\noaffiliation
\affiliation{University of the Basque Country UPV/EHU, 48080 Bilbao}
\affiliation{Budker Institute of Nuclear Physics SB RAS and Novosibirsk State University, Novosibirsk 630090}
\affiliation{Faculty of Mathematics and Physics, Charles University, 121 16 Prague}
\affiliation{Deutsches Elektronen--Synchrotron, 22607 Hamburg}
\affiliation{Justus-Liebig-Universit\"at Gie\ss{}en, 35392 Gie\ss{}en}
\affiliation{Gifu University, Gifu 501-1193}
\affiliation{The Graduate University for Advanced Studies, Hayama 240-0193}
\affiliation{Hanyang University, Seoul 133-791}
\affiliation{University of Hawaii, Honolulu, Hawaii 96822}
\affiliation{High Energy Accelerator Research Organization (KEK), Tsukuba 305-0801}
\affiliation{IKERBASQUE, Basque Foundation for Science, 48013 Bilbao}
\affiliation{Indian Institute of Technology Guwahati, Assam 781039}
\affiliation{Indian Institute of Technology Madras, Chennai 600036}
\affiliation{Institute of High Energy Physics, Chinese Academy of Sciences, Beijing 100049}
\affiliation{Institute of High Energy Physics, Vienna 1050}
\affiliation{Institute for High Energy Physics, Protvino 142281}
\affiliation{INFN - Sezione di Torino, 10125 Torino}
\affiliation{Institute for Theoretical and Experimental Physics, Moscow 117218}
\affiliation{J. Stefan Institute, 1000 Ljubljana}
\affiliation{Kanagawa University, Yokohama 221-8686}
\affiliation{Institut f\"ur Experimentelle Kernphysik, Karlsruher Institut f\"ur Technologie, 76131 Karlsruhe}
\affiliation{Kennesaw State University, Kennesaw GA 30144}
\affiliation{Department of Physics, Faculty of Science, King Abdulaziz University, Jeddah 21589}
\affiliation{Korea Institute of Science and Technology Information, Daejeon 305-806}
\affiliation{Korea University, Seoul 136-713}
\affiliation{Kyungpook National University, Daegu 702-701}
\affiliation{\'Ecole Polytechnique F\'ed\'erale de Lausanne (EPFL), Lausanne 1015}
\affiliation{Faculty of Mathematics and Physics, University of Ljubljana, 1000 Ljubljana}
\affiliation{University of Maribor, 2000 Maribor}
\affiliation{Max-Planck-Institut f\"ur Physik, 80805 M\"unchen}
\affiliation{School of Physics, University of Melbourne, Victoria 3010}
\affiliation{Moscow Physical Engineering Institute, Moscow 115409}
\affiliation{Moscow Institute of Physics and Technology, Moscow Region 141700}
\affiliation{Graduate School of Science, Nagoya University, Nagoya 464-8602}
\affiliation{Kobayashi-Maskawa Institute, Nagoya University, Nagoya 464-8602}
\affiliation{Nara Women's University, Nara 630-8506}
\affiliation{National Central University, Chung-li 32054}
\affiliation{National United University, Miao Li 36003}
\affiliation{Department of Physics, National Taiwan University, Taipei 10617}
\affiliation{H. Niewodniczanski Institute of Nuclear Physics, Krakow 31-342}
\affiliation{Niigata University, Niigata 950-2181}
\affiliation{Osaka City University, Osaka 558-8585}
\affiliation{Pacific Northwest National Laboratory, Richland, Washington 99352}
\affiliation{Peking University, Beijing 100871}
\affiliation{University of Pittsburgh, Pittsburgh, Pennsylvania 15260}
\affiliation{University of Science and Technology of China, Hefei 230026}
\affiliation{Soongsil University, Seoul 156-743}
\affiliation{Sungkyunkwan University, Suwon 440-746}
\affiliation{Department of Physics, Faculty of Science, University of Tabuk, Tabuk 71451}
\affiliation{Tata Institute of Fundamental Research, Mumbai 400005}
\affiliation{Excellence Cluster Universe, Technische Universit\"at M\"unchen, 85748 Garching}
\affiliation{Toho University, Funabashi 274-8510}
\affiliation{Tohoku University, Sendai 980-8578}
\affiliation{Department of Physics, University of Tokyo, Tokyo 113-0033}
\affiliation{Tokyo Institute of Technology, Tokyo 152-8550}
\affiliation{Tokyo Metropolitan University, Tokyo 192-0397}
\affiliation{University of Torino, 10124 Torino}
\affiliation{Utkal University, Bhubaneswar 751004}
\affiliation{CNP, Virginia Polytechnic Institute and State University, Blacksburg, Virginia 24061}
\affiliation{Wayne State University, Detroit, Michigan 48202}
\affiliation{Yamagata University, Yamagata 990-8560}
\affiliation{Yonsei University, Seoul 120-749}
  \author{Y.~M.~Goh}\affiliation{Hanyang University, Seoul 133-791} 
  \author{Y.~Unno}\affiliation{Hanyang University, Seoul 133-791} 
  \author{B.~G.~Cheon}\affiliation{Hanyang University, Seoul 133-791} 
  \author{A.~Abdesselam}\affiliation{Department of Physics, Faculty of Science, University of Tabuk, Tabuk 71451} 
  \author{I.~Adachi}\affiliation{High Energy Accelerator Research Organization (KEK), Tsukuba 305-0801}\affiliation{The Graduate University for Advanced Studies, Hayama 240-0193} 
  \author{H.~Aihara}\affiliation{Department of Physics, University of Tokyo, Tokyo 113-0033} 
  \author{S.~Al~Said}\affiliation{Department of Physics, Faculty of Science, University of Tabuk, Tabuk 71451}\affiliation{Department of Physics, Faculty of Science, King Abdulaziz University, Jeddah 21589} 
  \author{K.~Arinstein}\affiliation{Budker Institute of Nuclear Physics SB RAS and Novosibirsk State University, Novosibirsk 630090} 
 \author{D.~M.~Asner}\affiliation{Pacific Northwest National Laboratory, Richland, Washington 99352} 
  \author{V.~Aulchenko}\affiliation{Budker Institute of Nuclear Physics SB RAS and Novosibirsk State University, Novosibirsk 630090} 
% ...
\collaboration{The Belle Collaboration}

%% file: pub_arxiv.bbl
\begin{thebibliography}{99}
\bibitem{conj} Charge-conjugate processes are implied throughout the paper unless explicitly stated otherwise.
\end{thebibliography}
